%% file: bc0552_sub.tex
\documentclass[aps,prl,preprint,tightenlines,superscriptaddress]{revtex4}

\usepackage{graphicx} 
\usepackage{dcolumn}  

\graphicspath{{ps}}

\def\myspecial#1{}                   

\def\Journal#1#2#3#4{{#1} {\bf #2}, #3 (#4)}

\def\NIMA{Nucl. Instr. and Meth. A}

\def\PLB{{Phys. Lett.}  B}
\def\PRL{Phys. Rev. Lett.}
\def\PRD{{Phys. Rev.} D}

\def\be{\begin{equation}}
\def\ee{\end{equation}}
\def\bea{\begin{eqnarray}}
\def\eea{\end{eqnarray}}

\def\nbb{386 million $B\bar{B}$ pairs}
\def\dsup{\ensuremath{D_{\rm sup}}}
\def\dfav{\ensuremath{D_{\rm fav}}}
\def\rb{\ensuremath{r_B}}
\def\rd{\ensuremath{r_D}}
\def\deltab{\ensuremath{\delta_B}}
\def\deltad{\ensuremath{\delta_D}}
\def\stat{\ensuremath{\mathrm{stat}}}
\def\syst{\ensuremath{\mathrm{syst}}}
\def\LR{{\cal R}}

\def\RDK{R_{DK}}
\def\RDpi{R_{D\pi}}
\def\ADpi{{\cal A}_{D\pi}}

\def\DsupKyield{\ensuremath{2.4 \,^{+4.9}_{-4.4}}}
\def\DfavKyield{\ensuremath{634 \,^{+59}_{-99}}}
\def\Dsuppiyield{\ensuremath{50 \,^{+10}_{-11}}}
\def\Dfavpiyield{\ensuremath{14518 \pm 125}}
\def\DsupKsdbyield{\ensuremath{7.3 \,^{+6.9}_{-6.0}}}
\def\DsupKpkbgd{\ensuremath{2.4 \,^{+2.3}_{-2.0}}}
\def\DsupKcorryield{\ensuremath{0.0 \,^{+5.3}_{-5.0}}}

\def\BmDsuppiyield{\ensuremath{25.6 \,^{+7.4}_{-6.7}}}
\def\BpDsuppiyield{\ensuremath{21.0 \,^{+7.7}_{-7.0}}}

\def\RDKresult{\ensuremath{( 0.0 \,^{+8.4}_{-7.9} \, (\stat) \pm 1.0 \, (\syst) ) \times 10^{-3}}}
\def\RDpiresult{\ensuremath{( 3.5 \,^{+0.8}_{-0.7} \, (\stat) \pm 0.3 \,(\syst) ) \times 10^{-3}}}
\def\ADpiresult{\ensuremath{0.10 \pm 0.22 \,(\stat) \pm 0.02 \,(\syst)}}
\def\RDKlimit{\ensuremath{ 13.9 \times 10^{-3} }}
\def\rblimit{\ensuremath{0.18}}

\begin{document}

\preprint{
  \vbox{ 
    \hbox{   }
    \hbox{BELLE-CONF-0552}
    \hbox{EPS05-527} 
    \hbox{\today}
  }
}

\title{
  \quad\\[0.5cm]  
  \boldmath 
  Study of the Suppressed Decays 
  $B^\pm \to [K^\mp \pi^\pm]_D K^\pm$ and $B^\pm \to [K^\mp \pi^\pm]_D \pi^\pm$ at Belle
}

\input{author-conf2005}

\begin{abstract}
  We report an updated study of the suppressed decays 
  $B^- \to [K^+\pi^-]_D K^-$ and $B^- \to [K^+\pi^-]_D \pi^-$
  where $[K^+\pi^-]_D$ indicates that the $K^+\pi^-$ pair 
  originates from a neutral $D$ meson. 
  A data sample containing \nbb\ 
  recorded at the $\Upsilon(4S)$ resonance with the Belle detector 
  at the KEKB asymmetric $e^{+}e^{-}$ storage ring is used. 
  This decay mode is sensitive to the CKM angle $\phi_{3}$.
  We do not see a significant signal for $B^{-} \to [K^{+}\pi^{-}]_{D}K^{-}$,
  and we set a limit on the ratio of $B$ decay amplitudes $r_B < \rblimit$ 
  at the $90\%$ confidence level. 
  We measure the $CP$ asymmetry of the $B^- \to [K^+\pi^-]_D \pi^-$ mode, 
  $\ADpi = \ADpiresult$.
\end{abstract}

\maketitle

{\renewcommand{\thefootnote}{\fnsymbol{footnote}}}
\setcounter{footnote}{0}

\section{\boldmath Introduction}
\begin{figure}[h]
  \begin{center}
    \includegraphics[width=0.5\textwidth]{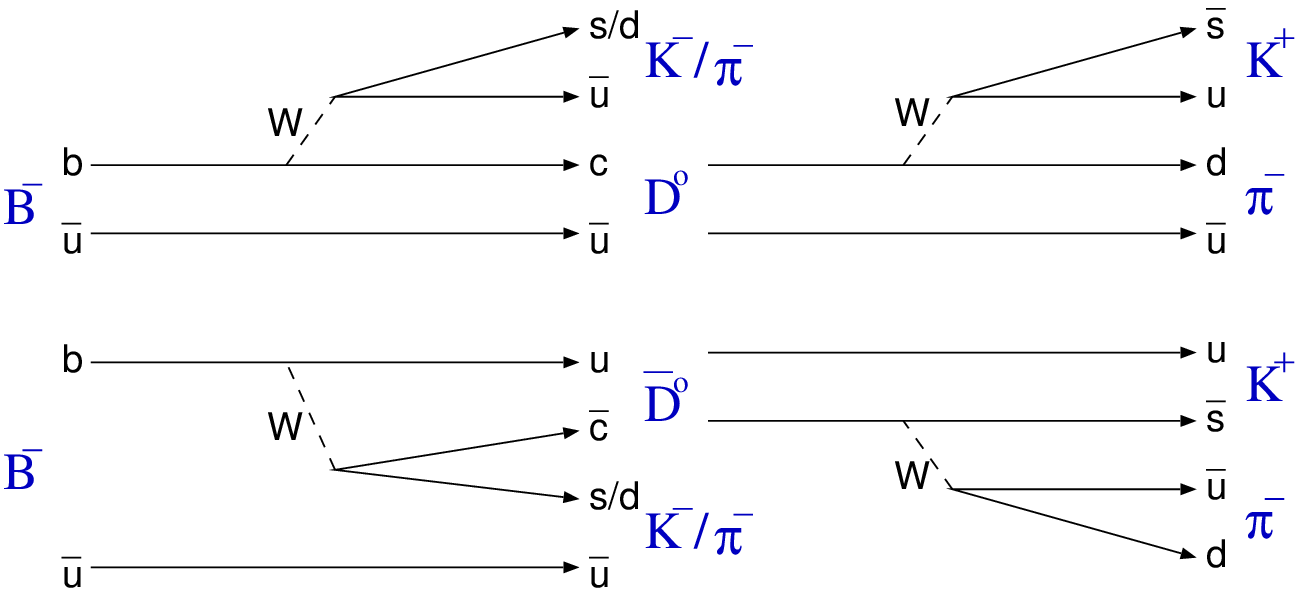}
    \caption{$B^- \to [K^+\pi^-]_D K^-$ and $B^- \to [K^+\pi^-]_D \pi^-$ decays.}
    \label{fig:btodcsk}
  \end{center}
\end{figure}

Precise measurements of the elements of the 
Cabibbo-Kobayashi-Maskawa matrix~\cite{km} 
constrain the Standard Model and may reveal new physics.
However, the extraction of the Unitarity Triangle angle $\phi_3$~\cite{ut}
is a challenging measurement even with modern high luminosity $B$ factories. 
Several methods for measuring $\phi_3$ use the interference 
between $B^- \to D^0 K^-$ and $B^- \to \bar{D}^0 K^-$, 
which occurs when $D^0$ and $\bar{D}^0$ decay
to common final states~\cite{bs,glw}.
$CP$ violation occurs when both weak and strong phase differences 
between the amplitudes are non-trivial.
As noted by Atwood, Dunietz and Soni (ADS)~\cite{ads},
$CP$ violation effects are enhanced if the final state is chosen so
that the interfering amplitudes have comparable magnitudes;
the archetype uses $B^- \to [K^+\pi^-]_D K^-$,
where $[K^+\pi^-]_{D}$ indicates that the $K^+\pi^-$ pair 
originates from a neutral $D$ meson. 
In this case, the colour-allowed $B$ decay 
followed by the doubly Cabibbo-suppressed $D$ decay 
interferes with  the colour-suppressed $B$ decay 
followed by the Cabibbo-allowed $D$ decay (Fig.~\ref{fig:btodcsk}). 
Previous studies of this decay mode 
by BaBar~\cite{babar} and Belle~\cite{saigo}
have not found any significant signals for $B^- \to [K^+\pi^-]_D K^-$.
For the suppressed decay $B^- \to [K^+\pi^-]_D \pi^-$,
both topology and phenomenology are similar to $B^- \to [K^+\pi^-]_D K^-$;
our previous publication reported 
the first observation of this mode~\cite{saigo}. 

\section{\boldmath Analysis}
In this paper, we report an updated analysis of the suppressed decays 
$B^\pm \to [K^\mp\pi^\pm]_D K^\pm$ and $B^\pm \to [K^\mp\pi^\pm]_D \pi^\pm$.
In addition, the allowed decays $B^\pm \to [K^\pm\pi^\mp]_D K^\pm$ and 
$B^\pm \to [K^\pm\pi^\mp]_D\pi^\pm$ are used as control samples 
to reduce systematic uncertainties. 
The same selection criteria for the suppressed decay modes 
are applied to the control samples whenever possible. 
The main changes with respect to our previous publication~\cite{saigo}
are the inclusion of additional data 
corresponding to 111 million $B\bar{B}$ pairs,
and improved suppression of the dominant continuum background.
Throughout this report, charge conjugate states are implied 
except where explicitly mentioned and 
we denote the analysed decay modes as follows:
\begin{eqnarray}
  & \textrm{Suppressed decay} & 
  B^- \to [K^+ \pi^-]_D h^-: \:\:\:\: B^- \to \dsup h^- \nonumber \\
  & \textrm{Allowed decay} & 
  B^- \to [K^- \pi^+]_D h^-: \:\:\:\: B^- \to \dfav h^- \:\:\: (h = K, \pi). \nonumber
\end{eqnarray}
The results are based on a data sample containing $386$ million 
$B\bar{B}$ pairs collected at the $\Upsilon(4S)$ resonance
with the Belle detector 
at the KEKB asymmetric energy $e^+ e^-$ collider~\cite{KEKB}.
The Belle detector is a large-solid-angle magnetic
spectrometer that consists of a silicon vertex detector (SVD),
a 50-layer central drift chamber (CDC), an array of
aerogel threshold \v{C}erenkov counters (ACC),
a barrel-like arrangement of time-of-flight
scintillation counters (TOF), 
and an electromagnetic calorimeter (ECL) comprised of CsI(Tl) crystals 
located inside a superconducting solenoid coil that provides a 
$1.5 \ {\rm T}$ magnetic field.  
An iron flux-return located outside of the coil is instrumented 
to detect $K_L^0$ mesons and to identify muons (KLM).  
The detector is described in detail elsewhere~\cite{Belle}.
Two different inner detector configurations were used. For the first sample
of 152 million $B\bar{B}$ pairs, a 2.0 cm radius beampipe
and a 3-layer silicon vertex detector were used;
for the latter 234 million $B\bar{B}$ pairs,
a 1.5 cm radius beampipe, a 4-layer silicon detector
and a small-cell inner drift chamber were used~\cite{Ushiroda}.

\subsection{\boldmath Event selection}
$D$ mesons are reconstructed by combining two oppositely charged tracks. 
These charged tracks are required to have a point of closest approach 
to the beam line within $\pm5 \ {\rm mm}$ of the interaction point 
in the direction perpendicular to the beam axis $(dr)$ and 
$\pm5 \ {\rm cm}$ in the direction antiparallel to the positron beam axis $(dz)$.
A $K/\pi$ likelihood ratio 
$P(K/\pi) = \mathcal{L}_{K}/(\mathcal{L}_{K} + \mathcal{L}_{\pi})$ 
is formed for each track, 
where $\mathcal{L}_{K}$ and $\mathcal{L}_{\pi}$ are kaon and pion likelihoods,
calculated using $dE/dx$ measurements from the CDC, 
\v{C}erenkov light yields in the ACC 
and timing information from the TOF. 
We used the particle identification requirement 
$P(K/\pi) > 0.4$ and $P(K/\pi) < 0.7$ for kaons and pions from 
$D \to K\pi$ decays, respectively. 
$D$ candidates are required to have an invariant mass 
within $\pm2.5\sigma$ of the nominal $D^0$ mass: 
$1.850 \ {\rm GeV}/c^2 < M(K\pi) < 1.879 \ {\rm GeV}/c^2$. 
To improve the momentum determination, 
tracks from the $D$ candidate are refitted 
according to the nominal $D^0$ mass hypothesis 
and the reconstructed vertex position (a mass-and-vertex-constrained fit).

$B$ mesons are reconstructed by combining $D$ candidates 
with primary charged hadron candidates. 
For the primary charged tracks, 
we require $P(K/\pi) > 0.6$ for the kaon in $B^- \to DK^-$
and $P(K/\pi) < 0.2$ for the pion in $B^- \to D\pi^-$. 
The signal is identified by two kinematic variables, 
the energy difference $\Delta E = E_D + E_{K^- (\pi^-)} - E_{\rm beam}$ 
and the beam-energy-constrained mass 
$M_{\rm bc} = \sqrt{E^2_{\rm beam} - (\vec{p}_D + \vec{p}_{K^- (\pi^-)})^2}$, 
where $E_D$ is the energy of the $D$ candidate, 
$E_{K^- (\pi^-)}$ is the energy of the $K^- (\pi^-)$ and 
$E_{\rm beam}$ is the beam energy, in the centre of mass (cm) frame. 
$\vec{p}_D$ and $\vec{p}_{K^- (\pi^-)}$ are 
the momenta of the $D$ and $K^- (\pi^-)$ in the cm frame. 
We define the signal region as 
$5.27 \ {\rm GeV}/c^2 < M_{\rm bc} < 5.29 \ {\rm GeV}/c^2$ and 
$-0.05 \ {\rm GeV} < \Delta E < 0.05 \ {\rm GeV}$. 
In the case of multiple candidates, 
which occurs in $1$--$2\%$ of events with at least one candidate,
we choose the best candidate on the basis of a $\chi^2$ 
determined from the difference between the measured and nominal values 
of $M_{\rm bc}$.  
  
\subsection{\boldmath $q\bar{q}$ continuum suppression}
To suppress the large background from the two-jet like 
$e^+e^- \to q\bar{q} \ (q = u, d, s, c)$ continuum processes, 
variables that characterise the event topology are used. 
We construct a Fisher discriminant~\cite{fisher} 
of modified Fox-Wolfram moments~\cite{sfw},
which we denote $SFW$.
The Fisher coefficients are optimized by maximising the 
separation between signal events and continuum events. 
Furthermore, $\cos\theta_{B}$, the angle in the cm system of the
$B$ flight direction with respect to the beam axis is also used 
to distinguish $B\bar{B}$ events from continuum events. 
These two independent variables, $SFW$ and $\cos\theta_{B}$, 
are combined to form a likelihood ratio ($\LR$),
\begin{eqnarray}
  \LR & = & 
  {\cal L}_{\rm sig}/({\cal L}_{\rm sig} + {\cal L}_{\rm cont}) 
  \nonumber \\
  {\cal L}_{\rm sig (cont)} & = & 
  {\cal L}_{\rm sig (cont)}^{SFW} \times {\cal L}_{\rm sig(cont)}^{\cos\theta_{B}}, 
  \nonumber
\end{eqnarray}
where ${\cal L}_{\rm sig}$ and ${\cal L}_{\rm cont}$ 
are likelihoods defined from $SFW$ and $\cos\theta_{B}$ 
distributions for signal and continuum backgrounds, respectively. 
We optimize the $\LR$ requirement by maximising $S/\sqrt{S + N}$, 
where $S$ and $N$ denote the 
expected number of signal and background events in the signal region. 
The signal expectations are calculated from our previous results~\cite{saigo},
the efficiency obtained from Monte Carlo simulation (given later),
and the number of $B\bar{B}$ pairs;
the background expectations are obtained using events 
in the $M_{\rm bc}$ sideband 
($5.240 \ {\rm GeV}/c^2 < M_{\rm bc} < 5.265 \ {\rm GeV}/c^2$),
with the extrapolation into the signal region based on Monte Carlo.
For $B^- \to \dsup K^- (\pi^-)$ we require $\LR > 0.90 \ (0.74)$, 
which retains $40.0\% \ (65.7\%)$ of the signal and 
removes $99.0\% \ (94.3\%)$ of the continuum background. 

\subsection{\boldmath Peaking backgrounds}
\begin{figure}[htb]
  \begin{center}
    \includegraphics[width=0.45\textwidth,height=0.3\textwidth]{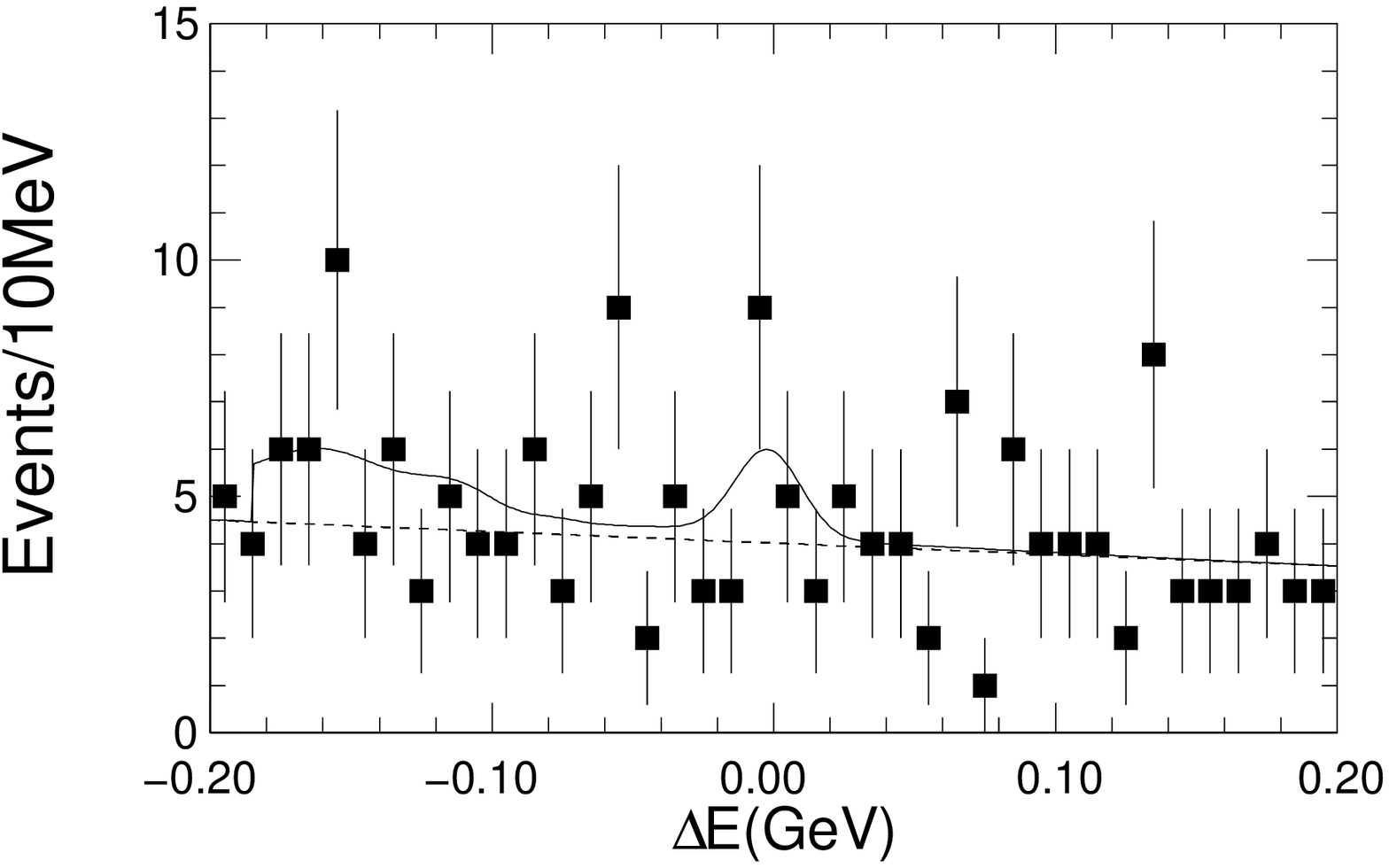}
    \includegraphics[width=0.45\textwidth,height=0.3\textwidth]{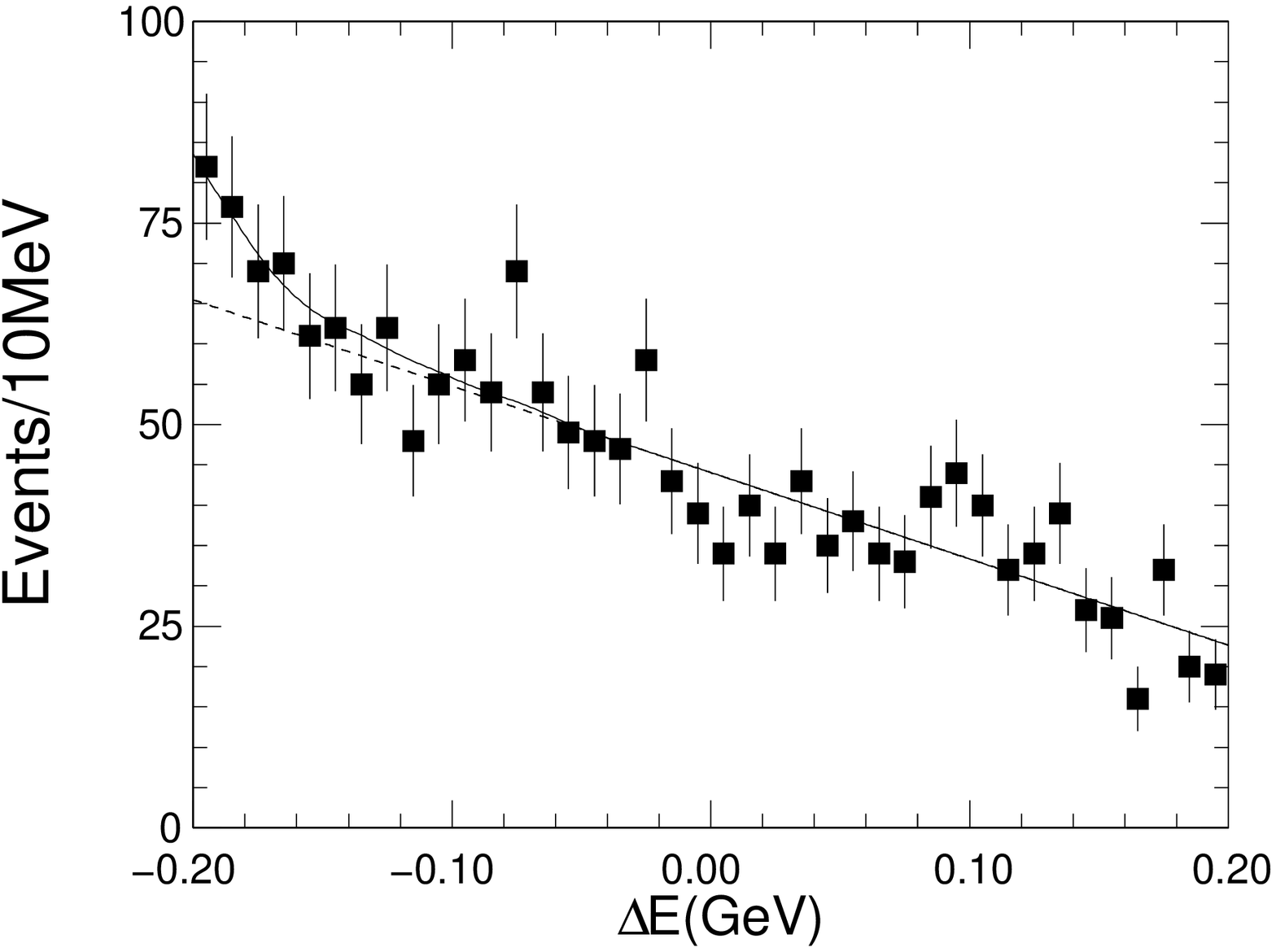}
    \caption{
      $\Delta E$ distributions for events in the $D^0$ mass sideband for 
      (left) $B^- \to \dsup K^-$ and (right) $B^- \to \dsup \pi^{-}$. 
      The signal shapes are modelled using the results of the 
      $B^- \to \dfav h^- \ (h=K,\pi)$ fit.
    }
    \label{fig:peakbkg}
  \end{center}
\end{figure}

For $B^- \to \dsup K^-$, 
one can have a contribution from $B^- \to D^0 \pi^-$, $D^0 \to K^+K^-$, 
which has the same final state and can peak under the signal. 
In order to reject these events, we veto events that satisfy 
$1.843 \ {\rm GeV}/c^2 < M(KK) < 1.894 \ {\rm GeV}/c^2$. 
The allowed decay $B^- \to \dfav h^-$ can also cause 
a peaking background for the suppressed decay modes 
due to $K\pi$ misidentification. 
Therefore, we veto events for which the invariant mass of the $K\pi$ pair 
is inside the $D$ mass cut window when the mass assignments are exchanged. 
Furthermore, three-body charmless decays 
$B^- \to K^+K^-\pi^-$ and $B^- \to K^+\pi^-\pi^-$ can peak inside the signal region for 
$B^- \to \dsup K^-$ and $B^- \to \dsup \pi^-$, respectively. 
These peaking backgrounds are estimated from the $\Delta E$ distributions 
of events in a $D$ mass sideband, corresponding to 
$\pm(2.5-10)\sigma$ away from the nominal $D$ mass
($1.807 \ {\rm GeV}/c^2 < M(K\pi) < 1.850 \ {\rm GeV}/c^2$ and 
$1.879 \ {\rm GeV}/c^2 < M(K\pi) < 1.937 \ {\rm GeV}/c^2$).
We fit these distributions,
which are shown in Fig.~\ref{fig:peakbkg},
using a procedure similar to that used for candidate signal events
(described later).
For $B^- \to \dsup \pi^-$, the peaking background estimated 
by fitting the plot is consistent with zero. 
Since the Standard Model prediction for the 
$B^- \to K^+\pi^-\pi^-$ branching fraction is smaller than $10^{-11}$~\cite{sm}, 
this background contribution is ignored. 
On the other hand, for $B^- \to \dsup K^-$, 
the peaking background yield in the $D$ mass sideband
is $\DsupKsdbyield$ events,
from which we expect $\DsupKpkbgd$ peaking background events
inside the $\Delta E$ signal region.

After applying all the cuts, 
the signal efficiencies are $14.6\%$ and $24.5\%$ for 
$B^- \to \dsup K^-$ and $B^- \to \dsup \pi^-$, respectively. 
The signal yields are extracted by fitting the $\Delta E$ distributions.

\subsection{\boldmath Fitting the $\Delta E$ distributions} 
Backgrounds from decays such as $B^- \to D\rho^-$ and $B^- \to D^*\pi^-$ 
are distributed in the negative $\Delta E$ region and make a small contribution 
to the signal region. 
The shape of this $B\bar{B}$ background is modelled as a 
smoothed histogram from generic Monte Carlo (MC) samples. 
The continuum background populates the entire $\Delta E$ region. 
We model its shape with as a first order polynomial. 
The signal $\Delta E$ distribution is modelled as the
sum of two Gaussian distributions with a common mean.

In the fit to the $\Delta E$ distribution of $B^- \to \dfav \pi^-$, 
the free parameters are the position, widths, area and 
fraction in the tail of the signal peak, 
the slope and normalisation of the continuum component
and the normalisation of the $B\bar{B}$ background.
For the $B^- \to \dfav K^-$ fit, 
there is an additional component due to feed-across from $\dfav \pi^-$, 
which we model with a Gaussian shape that has different widths 
on the left and right sides of the peak,
since the shift caused by wrong mass assignment makes the shape asymmetric. 
The normalisation and shape parameters of this function are 
free parameters of the fit 
(all parameters which are floated in the $B^- \to \dfav \pi^-$ fit
are again free parameters).

\begin{figure}[htb]
  \begin{center}
    \includegraphics[width=0.47\textwidth,height=0.35\textwidth]{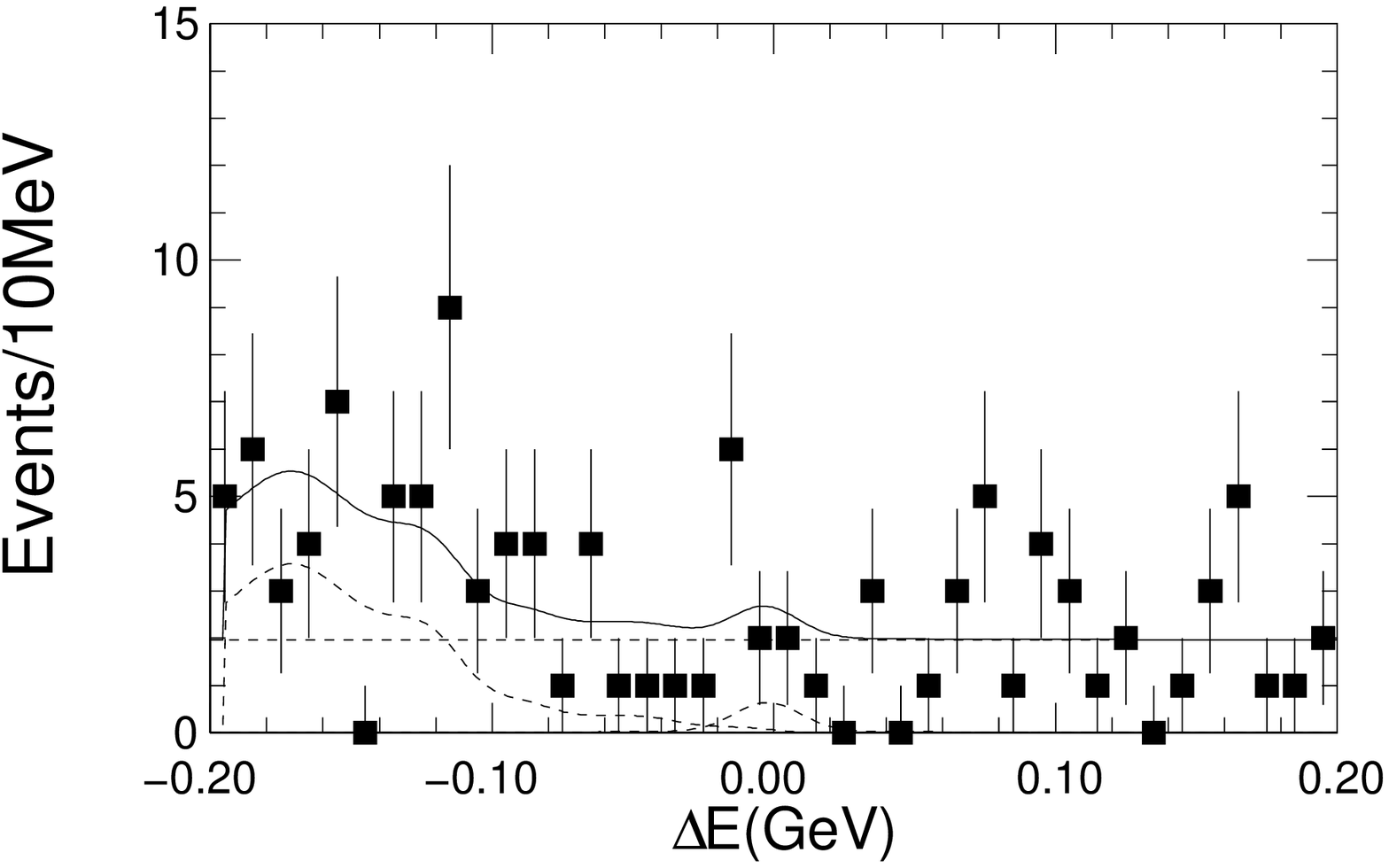}
    \includegraphics[width=0.47\textwidth,height=0.35\textwidth]{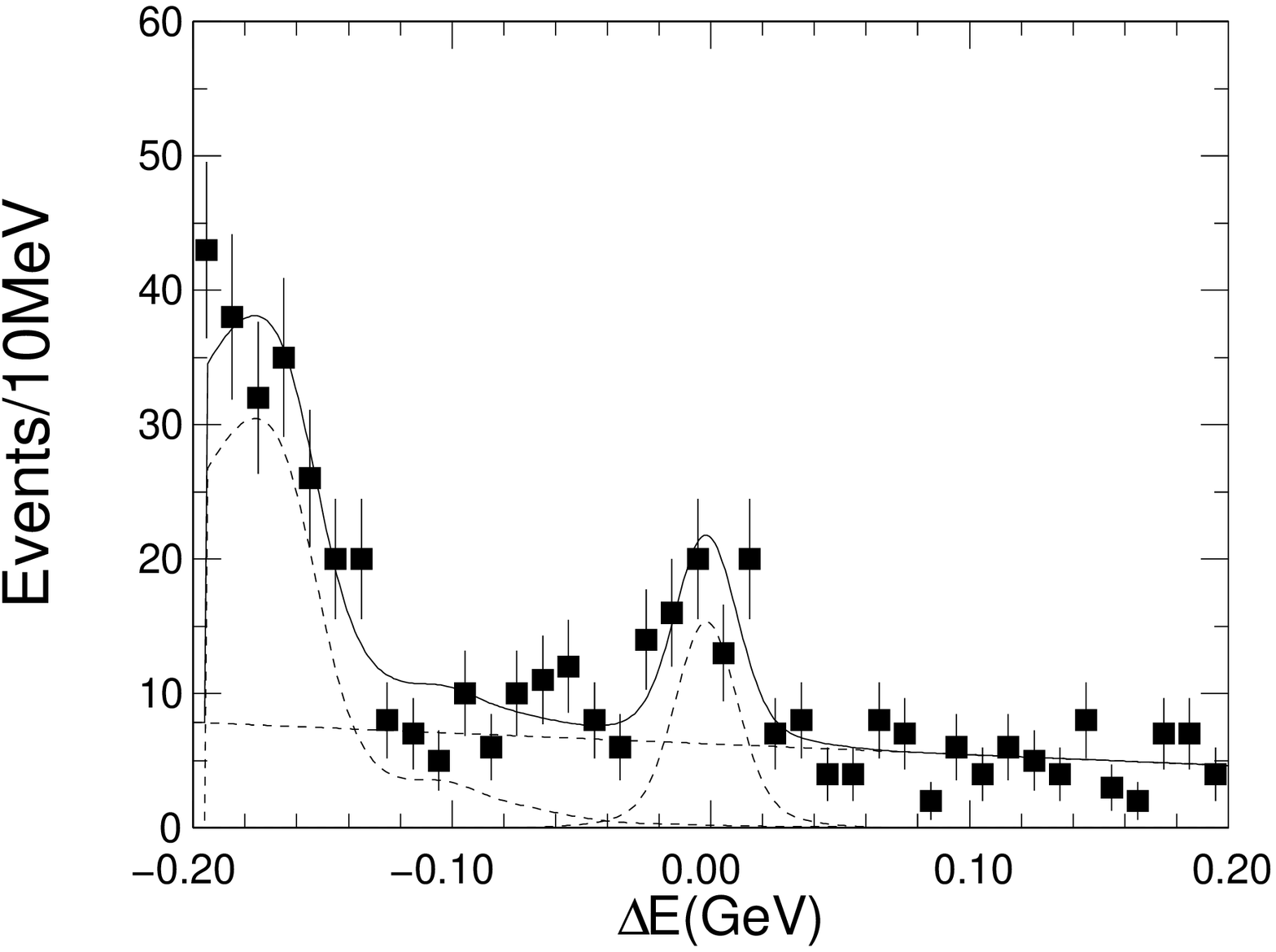}
    \includegraphics[width=0.47\textwidth,height=0.35\textwidth]{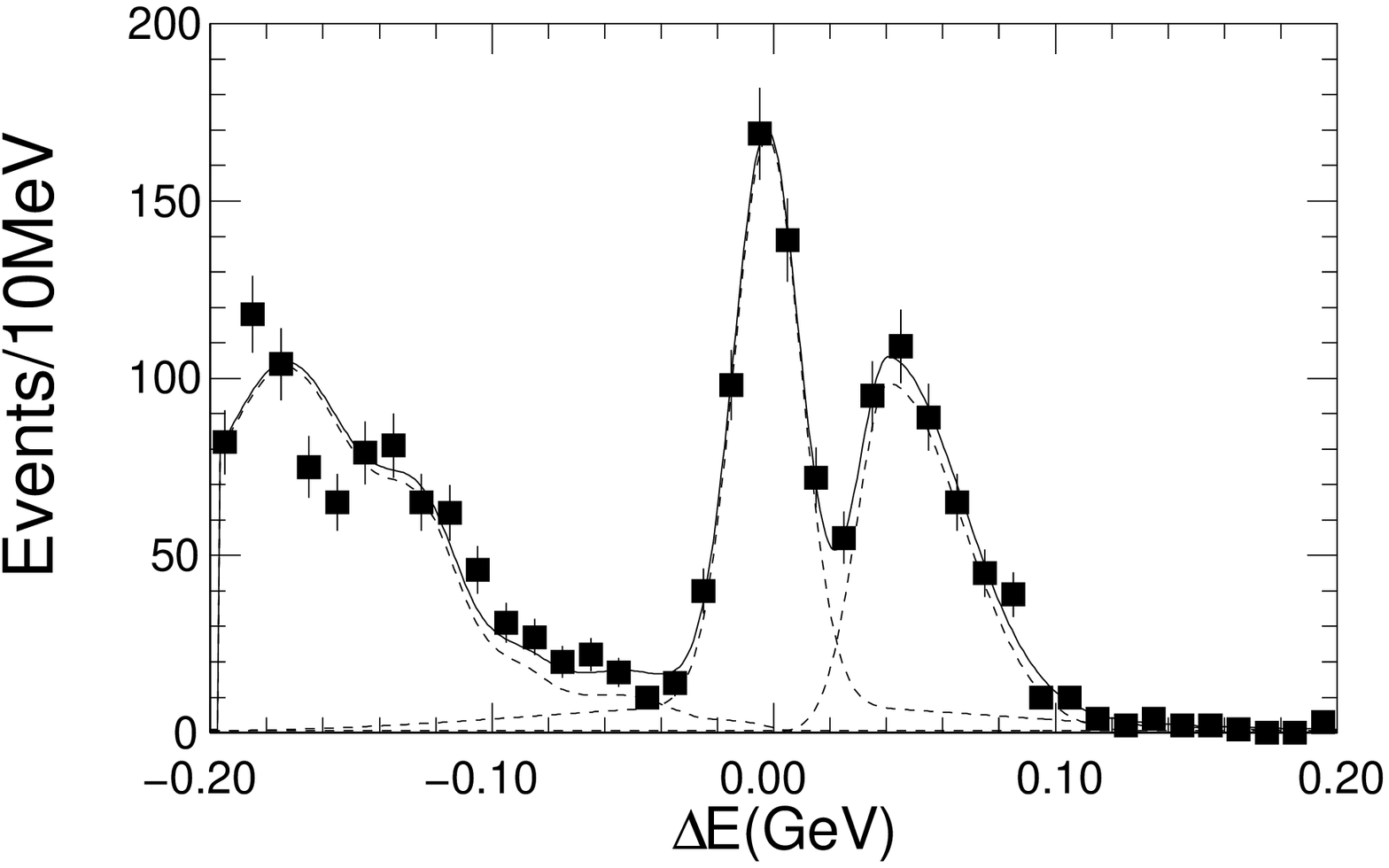}
    \includegraphics[width=0.47\textwidth,height=0.35\textwidth]{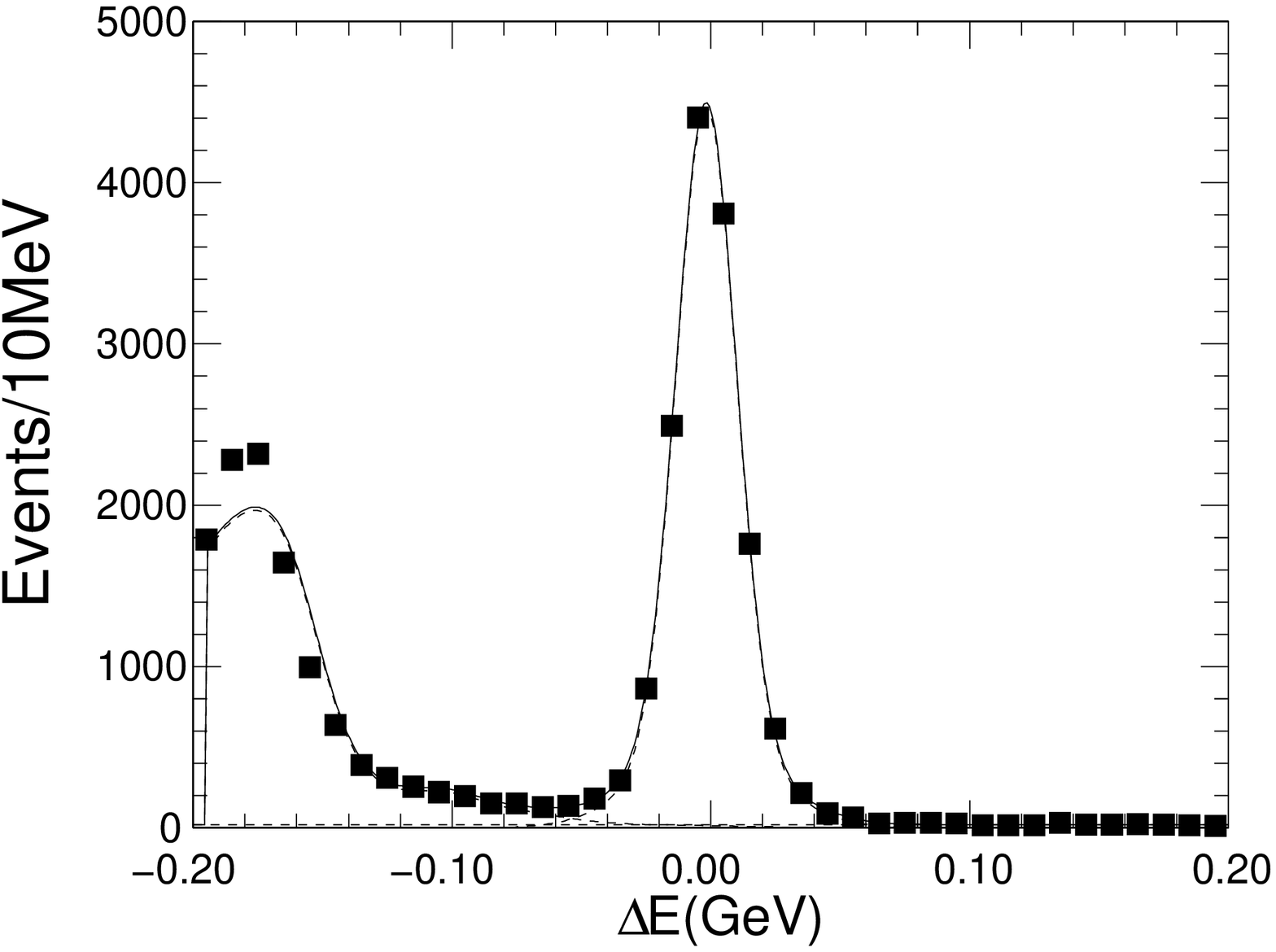}
    \caption{
      $\Delta E$ fit results for 
      (top left) $B^- \to \dsup K^-$, (top right) $B^- \to \dsup \pi^-$, 
      (bottom left) $B^- \to \dfav K^-$, (bottom right) $B^- \to \dfav \pi^-$.
      Charge conjugate modes are included.
    }
    \label{fig:fitting}
  \end{center}
\end{figure}

For $B^- \to \dsup K^-$ and $B^- \to \dsup \pi^-$, 
the signal and $B\bar{B}$ background shapes are modelled 
using the results of the fits to the corresponding favoured modes.
The free parameters are the normalisations of the three components,
and the slope of the continuum.
The amount of feed-across from $\dsup \pi^-$ to $\dsup K^-$
is a free parameter, but is found to be negligible, as expected.
The fit results are shown in Fig.~\ref{fig:fitting}. 
The numbers of events for $B^- \to \dsup h^-$ and $\dfav h^-$
are given in Table~\ref{tab:yield}.

\begin{table}
  \caption{
    Efficiencies and signal yields.
    For the $B^- \to \dsup K^-$ signal yield, 
    the second value is after subtraction of the peaking background.
  }
  \begin{tabular}{ l | c c }
    \hline
    \hline
    Mode & \hspace{3mm}Efficiency ($\%$)\hspace{3mm} & \hspace{3mm}Signal Yield\hspace{3mm} \\
    \hline
    $B^- \to \dsup K^-$ & $14.6 \pm 0.2$ & $\DsupKyield$ / $\DsupKcorryield$ \\
    $B^- \to \dsup \pi^-$ & $24.5 \pm 0.3$ & $\Dsuppiyield$ \\
    $B^- \to \dfav K^-$ & $14.5 \pm 0.2$ & $\DfavKyield$ \\
    $B^- \to \dfav \pi^-$ & $24.9 \pm 0.3$ & $\Dfavpiyield$  \\
    \hline
  \end{tabular}
  \label{tab:yield}
\end{table}

\section{\boldmath Results}

\subsection{\boldmath Ratio of branching fractions $R_{Dh}$}
We calculate ratios of product branching fractions, defined as
\begin{eqnarray}
  R_{Dh} & \equiv & 
  \frac{{\cal B}(B^- \to \dsup h^-)}{{\cal B}(B^- \to \dfav h^-)} = 
  \frac{N_{\dsup h^-}/\epsilon_{\dsup h^-}}{N_{\dfav h^-}/\epsilon_{\dfav h^-}} \ (h = K,\pi),
  \nonumber
\end{eqnarray}
where $N_{\dsup h}$ ($N_{\dfav h}$) and  $\epsilon_{\dsup h^-}$ ($\epsilon_{\dfav h^-}$)
are the number of signal events and the reconstruction efficiency for the decay
$B^- \to \dsup h^-$ ($B^- \to \dfav h^-$),
and are given in Table~\ref{tab:yield}.
We obtain
\begin{eqnarray}
  \RDK  & = & \RDKresult, \nonumber \\
  \RDpi & = & \RDpiresult. \nonumber
\end{eqnarray}
Since the signal for $B^- \to \dsup K^-$ is not significant,
we set an upper limit at the $90\%$ confidence level (C.L.) of 
$\RDK < \RDKlimit$,
where we take the likelihood function as a Gaussian distribution
with width given by the quadratic sum of statistical and systematic errors,
and the area is normalised in the physical region 
of positive branching fraction. 

Most of the systematic uncertainties from the detection efficiencies 
and the particle identification cancel when taking the ratios, 
since the kinematics of the $B^- \to \dsup h^-$ 
and $B^- \to \dfav h^-$ processes are similar. 
The systematic errors are due to 
uncertainties in the yield extraction
and the efficiency difference between 
$B^- \to \dsup h^-$ and $B^- \to \dfav h^-$. 
The uncertainties in the signal shapes and the $q\bar q$ background shapes 
are determined by varying the shape of the fitting function by $\pm1\sigma$. 
The uncertainties in the $B\bar B$ background shapes are determined 
by fitting the $\Delta E$ distribution in the region 
$-0.07 \ {\rm GeV} < \Delta E < 0.20 \ {\rm GeV}$ 
ignoring the $B\bar{B}$ background contributions \textemdash{}
this is the largest source of uncertainty: 
the signal yields are affected by $7.4\%$ and $28.4\%$ 
for $\dsup \pi$ and $\dsup K$ (before peaking background subtraction) 
respectively. 
The uncertainties in the efficiency differences are determined using signal MC.
The total systematic error is the sum in quadrature of the above uncertainties.

\begin{figure}[h]
  \begin{center}
    \includegraphics[width=0.7\textwidth]{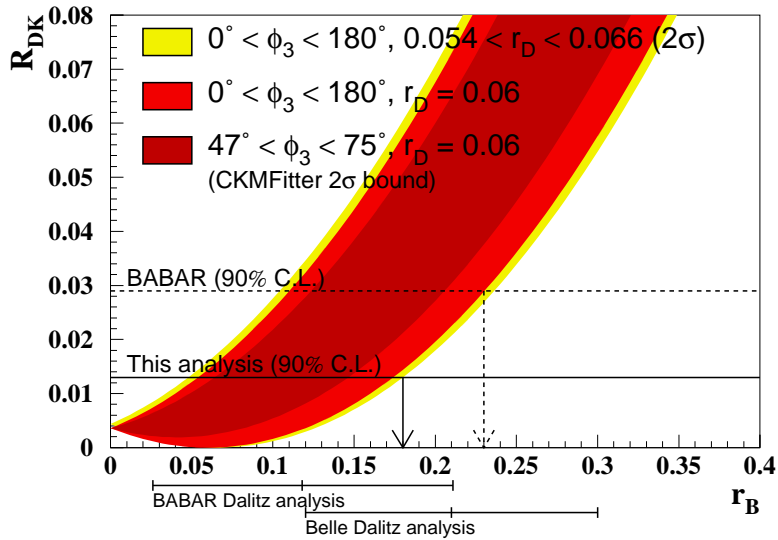}
    \caption{Constraint on $\rb$ from $\RDK$.}
    \label{fig:rb_rdk}
  \end{center}
\end{figure}

The ratio $\RDK$ is related to $\phi_3$ by
\begin{eqnarray}
  \RDK = \rb^2 + \rd^2 + 2 \rb \rd \cos \phi_3 \cos \delta, \nonumber
\end{eqnarray}
where~\cite{pdg}
\begin{eqnarray}
  \rb & \equiv & \left| \frac{A(B^- \to \bar{D}^0 K^-)}{A(B^- \to D^0 K^-)} \right|, 
  \:\:\:\:\: 
  \delta \equiv \deltab + \deltad, 
  \nonumber \\
  \rd & = & 
  \left|\frac{A(D^0 \to K^+\pi^-)}{A(D^0 \to K^-\pi^+)}\right| = 0.060 \pm 0.003.
  \nonumber
\end{eqnarray}
and $\deltab$ ($\deltad$) is the strong phase difference
between the two $B$ ($D$) decay amplitudes.
Using the above result, we obtain a limit on $\rb$. 
The least restrictive limit is obtained 
allowing $\pm 2\sigma$ variation on $\rd$ and assuming maximal interference
($\phi_3 = 0^\circ, \delta = 180^\circ$ or $\phi_3 = 180^\circ, \delta = 0^\circ$) 
and is found to be $\rb < \rblimit$ at the $90\%$ confidence level,
as shown in Fig.~\ref{fig:rb_rdk}.

\subsection{\boldmath $CP$ asymmetry}
\begin{figure}
  \begin{center}
    \includegraphics[width=0.47\textwidth]{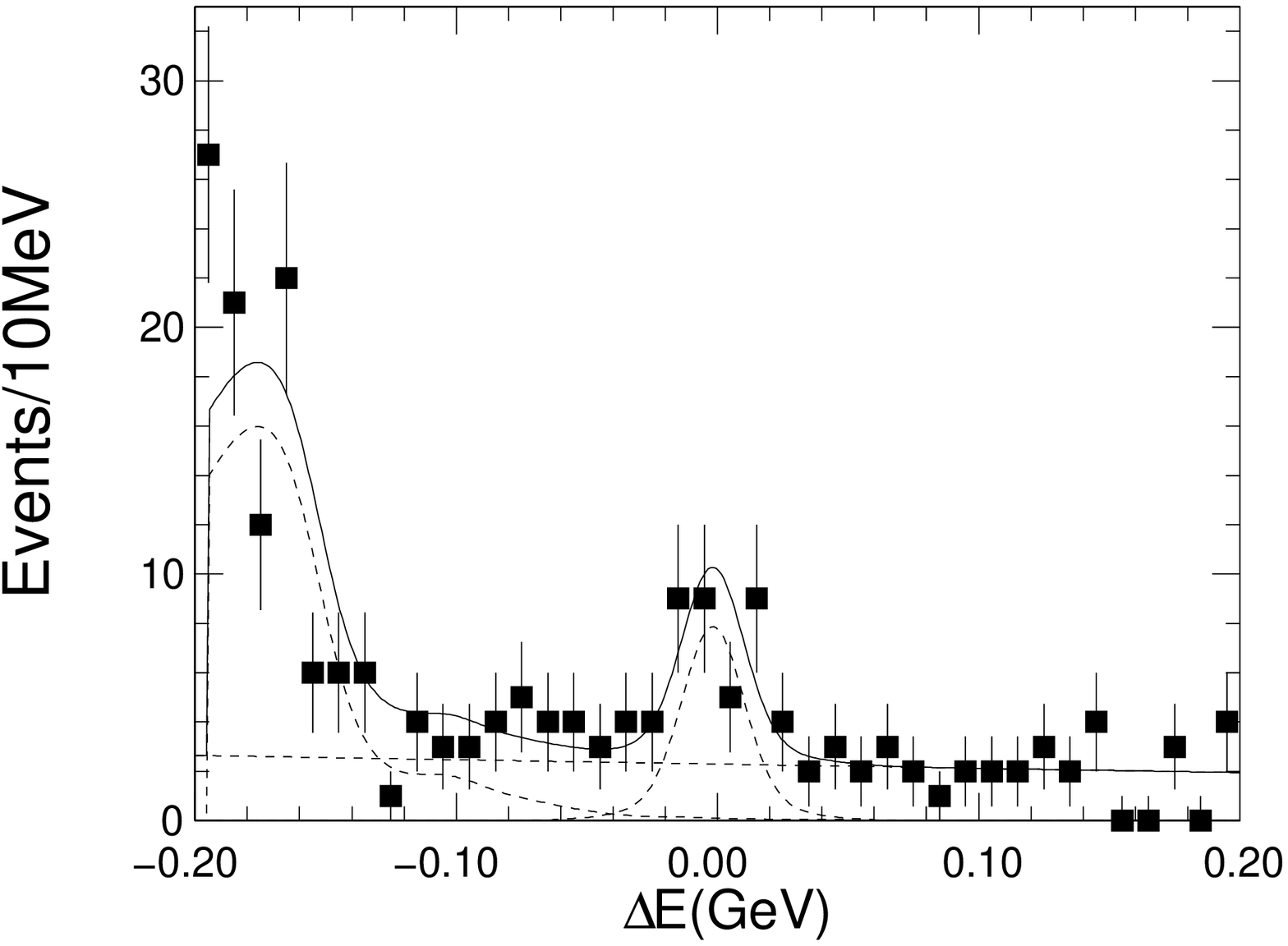}
    \includegraphics[width=0.47\textwidth]{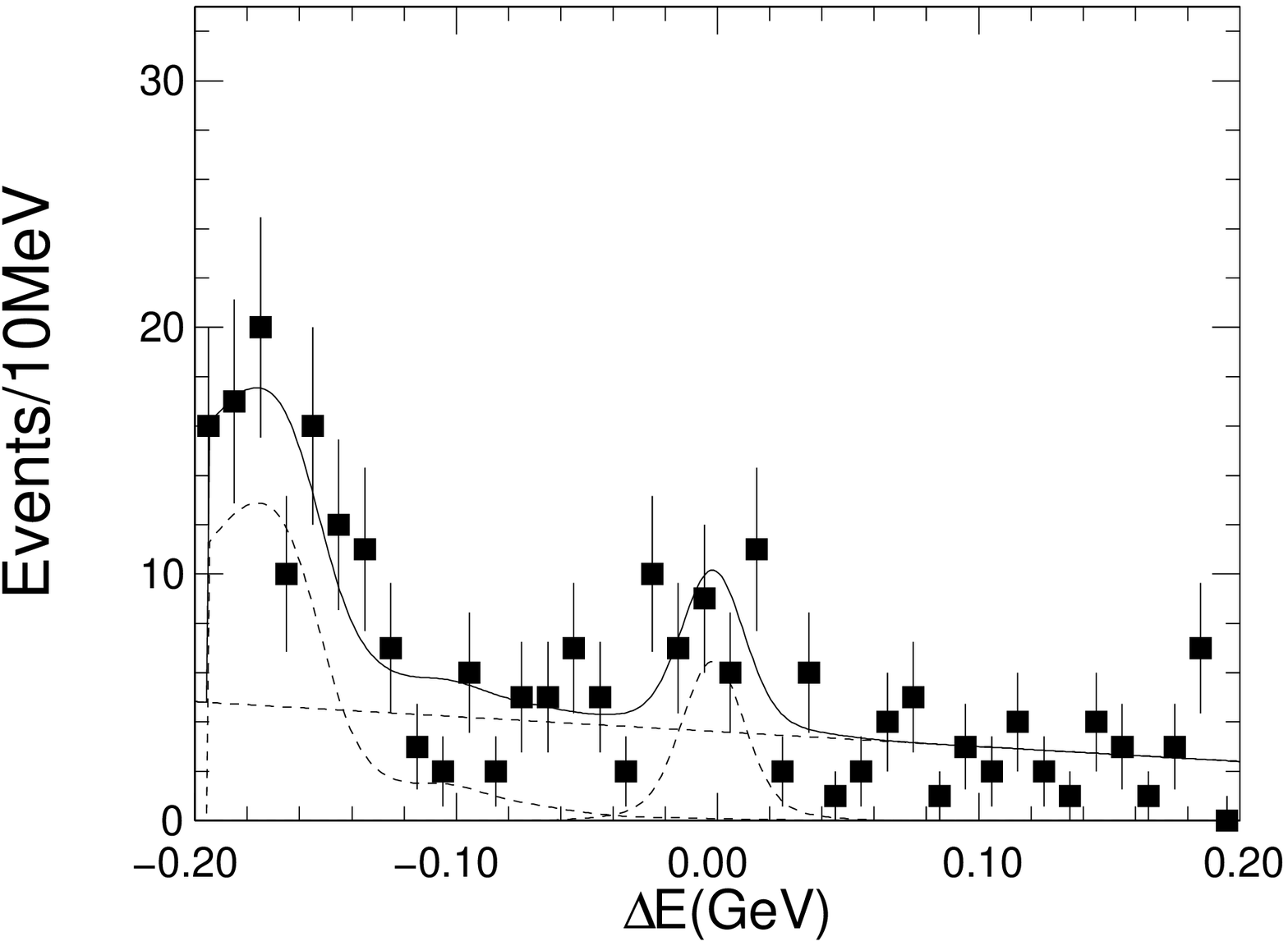}
    \caption{
      $\Delta E$ fit results for 
      (left) $B^- \to \dsup \pi^-$ and (right) $B^+ \to \dsup \pi^+$.
    }
    \label{fig:acp}
  \end{center}
\end{figure}

We search for $CP$ violating asymmetry in the $B^\pm \to \dsup \pi^\pm$ mode.
We fit the $B^+$ and $B^-$ yields separately, 
and determine $\ADpi$ as
\begin{eqnarray}
  \ADpi & \equiv & 
  \frac{
    {\cal B}(B^- \to \dsup \pi^-) - {\cal B}(B^+ \to \dsup \pi^+)
  }{
    {\cal B}(B^- \to \dsup \pi^-) + {\cal B}(B^+ \to \dsup \pi^+)
  }. \nonumber 
\end{eqnarray}
The fit results are shown in Fig.\ref{fig:acp}.
We find $\BmDsuppiyield$ $B^- \to \dsup \pi^-$ events and
$\BpDsuppiyield$ $B^+ \to \dsup \pi^+$ events,
giving an asymmetry of
\begin{eqnarray}
  \ADpi & = & \ADpiresult, \nonumber 
\end{eqnarray}
where systematic uncertainties arise from possible detector charge asymmetry 
($0.017$; determined from the $B^\pm \to \dfav \pi^\pm$ sample~\cite{acp_control}), 
and the $B^{+}$ and $B^{-}$ yield extraction 
($0.016$; determined as for $\RDpi$).
The total systematic error is obtained by taking the quadratic sum.
The measured partial rate asymmetry $\ADpi$ is consistent with zero. 

\section{\boldmath Summary}
Using \nbb\ collected with the Belle detector, 
we report studies of the suppressed decays $B^- \to \dsup h^-$ ($h=K,\pi$). 
We do not observe a signal for $B^- \to \dsup K^-$, 
and place a limit on the ratio of $B$ decay amplitudes $\rb < \rblimit$ 
at the $90\%$ confidence level. 
We have measured the $CP$ asymmetry in the related mode,
$\ADpi = \ADpiresult$.

\section*{Acknowledgements}
We thank the KEKB group for the excellent operation of the
accelerator, the KEK cryogenics group for the efficient
operation of the solenoid, and the KEK computer group and
the National Institute of Informatics for valuable computing
and Super-SINET network support. We acknowledge support from
the Ministry of Education, Culture, Sports, Science, and
Technology of Japan and the Japan Society for the Promotion
of Science; the Australian Research Council and the
Australian Department of Education, Science and Training;
the National Science Foundation of China under contract
No.~10175071; the Department of Science and Technology of
India; the BK21 program of the Ministry of Education of
Korea and the CHEP SRC program of the Korea Science and
Engineering Foundation; the Polish State Committee for
Scientific Research under contract No.~2P03B 01324; the
Ministry of Science and Technology of the Russian
Federation; the Ministry of Higher Education, 
Science and Technology of the Republic of Slovenia;  
the Swiss National Science Foundation; the National Science Council and
the Ministry of Education of Taiwan; and the U.S.\
Department of Energy.

\end{document}

%% file: author-conf2005.tex
\affiliation{Aomori University, Aomori}
\affiliation{Budker Institute of Nuclear Physics, Novosibirsk}
\affiliation{Chiba University, Chiba}
\affiliation{Chonnam National University, Kwangju}
\affiliation{University of Cincinnati, Cincinnati, Ohio 45221}
\affiliation{University of Frankfurt, Frankfurt}
\affiliation{Gyeongsang National University, Chinju}
\affiliation{University of Hawaii, Honolulu, Hawaii 96822}
\affiliation{High Energy Accelerator Research Organization (KEK), Tsukuba}
\affiliation{Hiroshima Institute of Technology, Hiroshima}
\affiliation{Institute of High Energy Physics, Chinese Academy of Sciences, Beijing}
\affiliation{Institute of High Energy Physics, Vienna}
\affiliation{Institute for Theoretical and Experimental Physics, Moscow}
\affiliation{J. Stefan Institute, Ljubljana}
\affiliation{Kanagawa University, Yokohama}
\affiliation{Korea University, Seoul}
\affiliation{Kyoto University, Kyoto}
\affiliation{Kyungpook National University, Taegu}
\affiliation{Swiss Federal Institute of Technology of Lausanne, EPFL, Lausanne}
\affiliation{University of Ljubljana, Ljubljana}
\affiliation{University of Maribor, Maribor}
\affiliation{University of Melbourne, Victoria}
\affiliation{Nagoya University, Nagoya}
\affiliation{Nara Women's University, Nara}
\affiliation{National Central University, Chung-li}
\affiliation{National Kaohsiung Normal University, Kaohsiung}
\affiliation{National United University, Miao Li}
\affiliation{Department of Physics, National Taiwan University, Taipei}
\affiliation{H. Niewodniczanski Institute of Nuclear Physics, Krakow}
\affiliation{Nippon Dental University, Niigata}
\affiliation{Niigata University, Niigata}
\affiliation{Nova Gorica Polytechnic, Nova Gorica}
\affiliation{Osaka City University, Osaka}
\affiliation{Osaka University, Osaka}
\affiliation{Panjab University, Chandigarh}
\affiliation{Peking University, Beijing}
\affiliation{Princeton University, Princeton, New Jersey 08544}
\affiliation{RIKEN BNL Research Center, Upton, New York 11973}
\affiliation{Saga University, Saga}
\affiliation{University of Science and Technology of China, Hefei}
\affiliation{Seoul National University, Seoul}
\affiliation{Shinshu University, Nagano}
\affiliation{Sungkyunkwan University, Suwon}
\affiliation{University of Sydney, Sydney NSW}
\affiliation{Tata Institute of Fundamental Research, Bombay}
\affiliation{Toho University, Funabashi}
\affiliation{Tohoku Gakuin University, Tagajo}
\affiliation{Tohoku University, Sendai}
\affiliation{Department of Physics, University of Tokyo, Tokyo}
\affiliation{Tokyo Institute of Technology, Tokyo}
\affiliation{Tokyo Metropolitan University, Tokyo}
\affiliation{Tokyo University of Agriculture and Technology, Tokyo}
\affiliation{Toyama National College of Maritime Technology, Toyama}
\affiliation{University of Tsukuba, Tsukuba}
\affiliation{Utkal University, Bhubaneswer}
\affiliation{Virginia Polytechnic Institute and State University, Blacksburg, Virginia 24061}
\affiliation{Yonsei University, Seoul}
  \author{K.~Abe}\affiliation{High Energy Accelerator Research Organization (KEK), Tsukuba} 
  \author{K.~Abe}\affiliation{Tohoku Gakuin University, Tagajo} 
  \author{I.~Adachi}\affiliation{High Energy Accelerator Research Organization (KEK), Tsukuba} 
  \author{H.~Aihara}\affiliation{Department of Physics, University of Tokyo, Tokyo} 
  \author{K.~Aoki}\affiliation{Nagoya University, Nagoya} 
  \author{K.~Arinstein}\affiliation{Budker Institute of Nuclear Physics, Novosibirsk} 
  \author{Y.~Asano}\affiliation{University of Tsukuba, Tsukuba} 
  \author{T.~Aso}\affiliation{Toyama National College of Maritime Technology, Toyama} 
  \author{V.~Aulchenko}\affiliation{Budker Institute of Nuclear Physics, Novosibirsk} 
  \author{T.~Aushev}\affiliation{Institute for Theoretical and Experimental Physics, Moscow} 
% ...
\collaboration{The Belle Collaboration}
\noaffiliation

%% file: bc0552_sub.bbl
\begin{thebibliography}{99}
\bibitem{acp_control}
  In the $B^\pm \to \dfav \pi^\pm$ sample we measure $\ADpi = -0.009 \pm 0.008$,
  from which we estimate a systematic error of $\pm 0.017$.
\end{thebibliography}
